\documentclass{article}
\usepackage{spconf,amsmath,graphicx}
\usepackage{romannum}
\usepackage[symbol]{footmisc}
\usepackage{hyperref}
\usepackage{subcaption}
\usepackage{multirow}
\usepackage{amsfonts}
\usepackage{bbm}
\usepackage{tabularx}

\title{MHTTS: FAST MULTI-HEAD TEXT-TO-SPEECH FOR SPONTANEOUS SPEECH WITH IMPERFECT TRANSCRIPTION}

\name{Dabiao Ma, Yitong Zhang, Meng Li, Feng Ye$^*$}
\address{360 Digital Technology, Beijing, China \\
\{madabiao-jk, zhangyitong-jk, limeng1-jk, yefeng\}@360shuke.com}
 
\begin{document}
\maketitle
\def\thefootnote{*}\footnotetext{Corresponding author.}
\def\thefootnote{$\dagger$}\footnotetext{Equal contribution.}

\begin{abstract}

Neural network based end-to-end Text-to-Speech (TTS) has greatly improved the quality of synthesized speech. While how to use massive spontaneous speech without transcription efficiently still remains an open problem. In this paper we propose MHTTS, a fast multi-speaker TTS system that is robust to transcription errors and speaking style speech data. Specifically, we introduce a multi-head model and transfer text information from high-quality corpus with manual transcription to spontaneous speech with imperfectly recognized transcription by jointly training them. MHTTS has three advantages: 1) Our system synthesizes better quality multi-speaker voice with faster inference speed. 2) Our system is capable of transferring correct text information to data with imperfect transcription, simulated using corruption, or provided by an Automatic Speech Recogniser
(ASR). 3) Our system can utilize massive real spontaneous speech with imperfect transcription and synthesize expressive voice.

\end{abstract}
\section{Introduction}
\label{sec:intro}

Text-to-speech(TTS) aims to synthesize natural and intelligible voice from text. Deep neutral networks\cite{deep_voice2, deep_voice3, tacotron2, transformertts, fastspeech, fastspeech2} have greatly improved the quality of synthesized speeches and expanded the capability of TTS system. And most TTS models can be trained only if recorded data with manual transcription is given. This kind of data is hard to collect while massive spontaneous speech without transcription is available in real life. How to utilize speech data with no transcription is becoming an important topic.

Adaptive methods\cite{ADASPEECH2, NAUTILUS} adapts on untranscribed speech data by fine-tuning a pre-trained multi-speaker TTS model but only takes dozens or hundreds of utterances for adaptation, which may not be able to model the speaking style of the spontaneous speech in real life. Another possible method is to generate transcripts by an ASR system\cite{kaldi, ContextNet, wenet} but the generated transcripts contain inevitable textual errors.
\cite{investigating_robustness} analyses impacts of different types of textual errors and concludes that attention-based autoregressive models\cite{tacotron2, DCTTS} are only partially robust to imperfect transcripts. Substitution and deletion errors pose a serious problem to autoregressive models.

Transformer\cite{transformer} based non-autoregressive models\cite{fastspeech, fastspeech2} greatly speedup synthesis process, synthesize more robust voice than autoregressive models and has reached state-of-the-art performance. While our experiments reveal that textual errors simulated or provided by ASR still highly affect the quality and pronunciation error rate of the synthesized voice.

Speech disentanglement\cite{Emergence_of_invariance, Unsupervised_Speech_Decomposition, Disentangling_Style_Factors} is a developing advanced topic that aims to decompose speech information into four disentangled components: text content, timbre, pitch, and rhythm. Timbre carries information of speaker's identity, while pitch and rhythm are related to the speaking style. Inspired by this perspective, if we can decompose the overall system into two disentangled sub-parts: one processing text information, the other merging text and speaker specific information (identity and speaking style), then we can design a proper training strategy to transfer text information of a large data corpus with correct transcription to the target spontaneous speech with imperfectly recognized transcription, without disturbing the learning of speaking style of the target spontaneous speech.

In this paper we propose MHTTS, a multi-speaker TTS model based on multi-head structure that has three main advantages:

\begin{itemize}

\item In common multi-speaker scenario, MHTTS speeds up the synthesis process and produces higher quality voice than transformer based non-autoregressive models, e.g. Fastspeech. Computation complexity of transformer is quadratic to the length of sequence, while computation complexity of MHTTS is only linear.
\item In two-speaker scenario where the transcription of one corpus is correct and that of the other is imperfect, MHTTS greatly reduces pronunciation errors of synthesized voice of the imperfect one by transferring text information.

\item MHTTS utilizes massive spontaneous speech in real life with transcription from ASR and synthesize spontaneous style voice with few pronunciation errors. In MHTTS, we try to disentangle the learning of text hidden representation and the learning of speaking style.


\end{itemize}

\section{METHODOLOGY}
\label{sec:Methodology}

\subsection{theory}
\label{subsec:theory}

\begin{figure}[h]
  \centering
  \centerline{\includegraphics[scale=0.7]{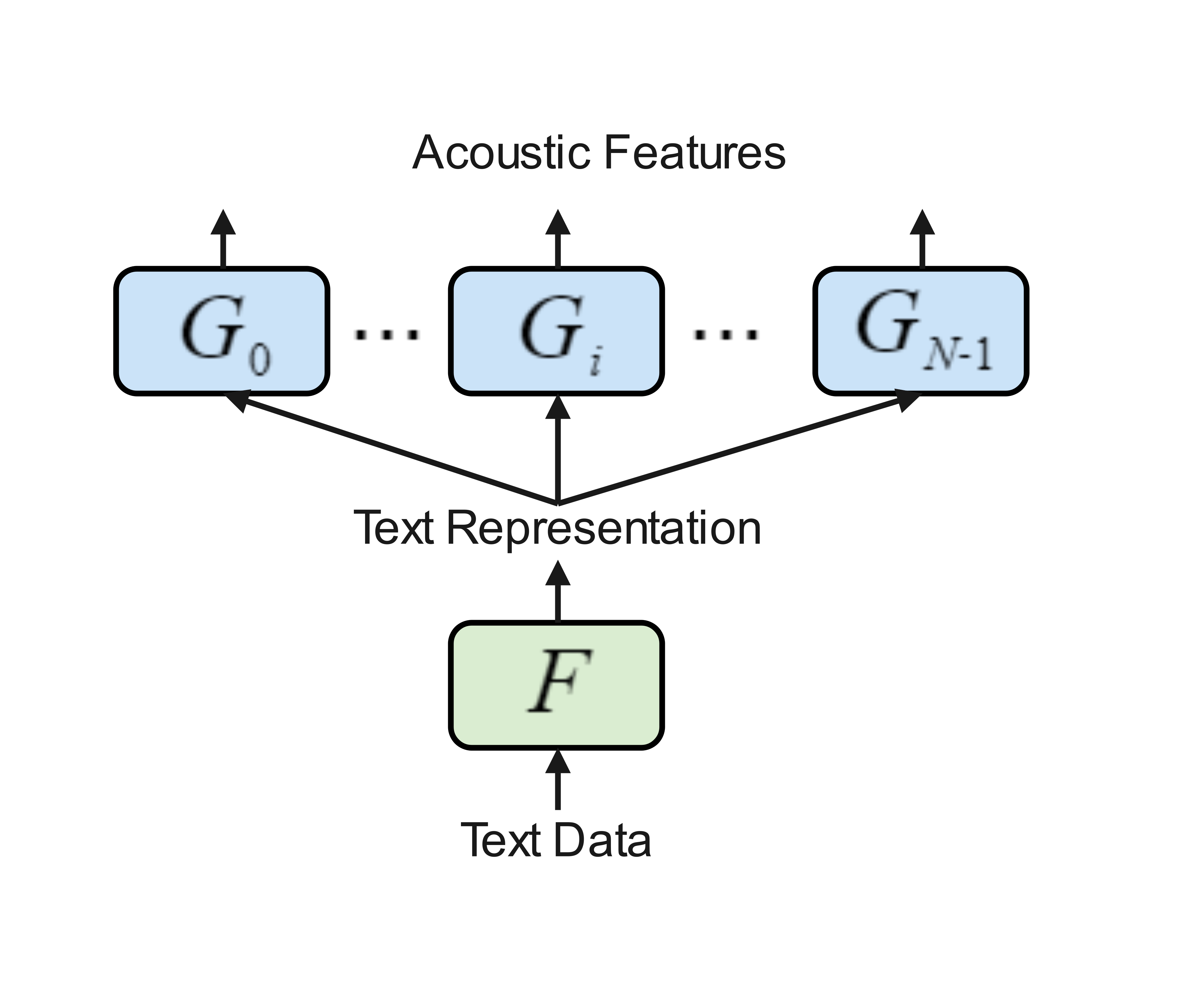}}
  \caption{Brief structure of our model. $F$ denotes the general text encoder block, $G$ denotes the speaker specific blocks.}\medskip
  \label{fig:fig1}
\end{figure}

Suppose there are corpora $D_{i, i=0,1,...,N-1}$ of $N$ different speakers, each $D_i$ contains text-speech pairs $(T_i, A_i)_{j, j = 0,...,M_i}$. Here $A$ denotes the acoustic feature that can be converted to speech by a neural vocoder\cite{wavenet, wavernn, melgan}. The following process is designed to predict $A_{i,j}$ from $T_{i,j}$: $T_{i,j}$ is first processed to get a text hidden representation $h_{i,j}$ by a general text encoder block $F$ which takes text data of all corpora as input, then $h_{i,j}$ is processed by a speaker specific block $G_i$ to predict $A_{i,j}$. As is shown by Fig.~\ref{fig:fig1}. $G_{i, i=0,1,...,N-1}$ forms the heads of MHTTS.



To disentangle the learning of text representation and the speaker related features, the model should satisfy the following conditions:

\begin{enumerate}

\item Text representations $h$ contains minimal speaker specific information.

\item Text representations $h$ should contain as much information about $A$ as possible, if the first condition is satisfied.

\item Speaker specific blocks $G$ has the capacity to reconstruct $A$ from $h$.

\end{enumerate}

The first condition guarantees that the weights of block $F$ contain minimal speaker specific information\cite{Emergence_of_invariance}. The first and second condition guarantee that block $F$ maps text data to good text hidden representations in the sense that $h$ can be easily converted to $A$ by speaker specific information contained in weights of block $G$. The second condition also helps the transfer of text information during training since all corpora share block $F$. Consider the extreme case that block $F$ is identity mapping, the multi-speaker TTS task degrades to $N$ disjoint single-speaker TTS tasks. The third condition is trivial.

From the perspective of the information bottleneck theory\cite{information_bottleneck}, if we use $I$ to denote the calculation of mutual information, these conditions can be expressed as maximizing the following quantity: 

\begin{align}
L=\frac{1}{N}\sum_i -\lambda I(h_i;\mathbbm{1})+\gamma I(h_i;A_i) + I(G_i(h_i);A_i)
\label{eqn:1}
\end{align}
where $\mathbbm{1}$ denotes the indicator vector of the speaker, $\lambda$ and $\gamma$ are weights. The terms in Eqn.\ref{eqn:1} correspond to the conditions respectively.

The maximization of the first term $-I(h_i;\mathbbm{1})$ is usually achieved by using gradient reversal layer (GRL)\cite{Domain_Adversarial} and a domain classifier\cite{Domain_Adversarial} to encourage $h_i$ to be speaker-invariant. But the performance of GRL is very sensitive to the hyperparameter and it is not applicable
in 2-speaker scenario especially when one corpus is much smaller. Instead we found that Layer Normalization (LN)\cite{Layer_Normalization} which normalizes the output distributions of block $F$ serves the same purpose well.

The maximization of the third term $I(G_i(h_i);A_i)$ is actually the minimization of the target cost function, e.g. the $L1$ or $L2$-norm distance between predicted acoustic features and the real features.

Now we consider the transformation $T_i \xrightarrow{F} h_i \xrightarrow{G_i} A_i$ to analyze the second term $I(h_i;A_i)$. Suppose in expectation $A_i$ is well constructed from $T_i$, that is $I(G_i(h_i);A_i)$ reaches a high value. Consider the case that $G_i$ is identity mapping, then $h_i = G_i(h_i)$ and  $I(h_i;A_i) = I(G_i(h_i);A_i)$ which is discouraged by the first term $-I(h_i;\mathbbm{1})$ since $A_i$ contains full information about $\mathbbm{1}$; If $G_i$ has a very large capacity, $F$ has the potential to degrade to identity mapping in which case $h_i = T_i$ and $I(h_i;A_i)$ reaches the minimal value $I(T_i;A_i)$ by the data processing inequality. In general, $G_i$ should have just sufficient capacity to contain speaker specific information and add the information to the text hidden representation $h_i$.


\subsection{architecture}

\subsubsection{block $F$}

The block $F$ transforms text data to text hidden representation  $h$. It is shared by all speaker corpora and occupies most of the computation complexity. 

Fastspeech\cite{fastspeech, fastspeech2} use Feed-Forward Transformer\cite{transformer} (FFT) blocks to build encoder and decoder. The multi-head attention structure\cite{transformer} in FFT makes it capable of modeling long-term dependency and performing parallel generation but also makes the computation complexity quadratic to the length of sequence.

\begin{figure}[h!]
  \begin{subfigure}[h]{0.5\textwidth}
  \centering
            \centerline{\includegraphics[scale=0.5]{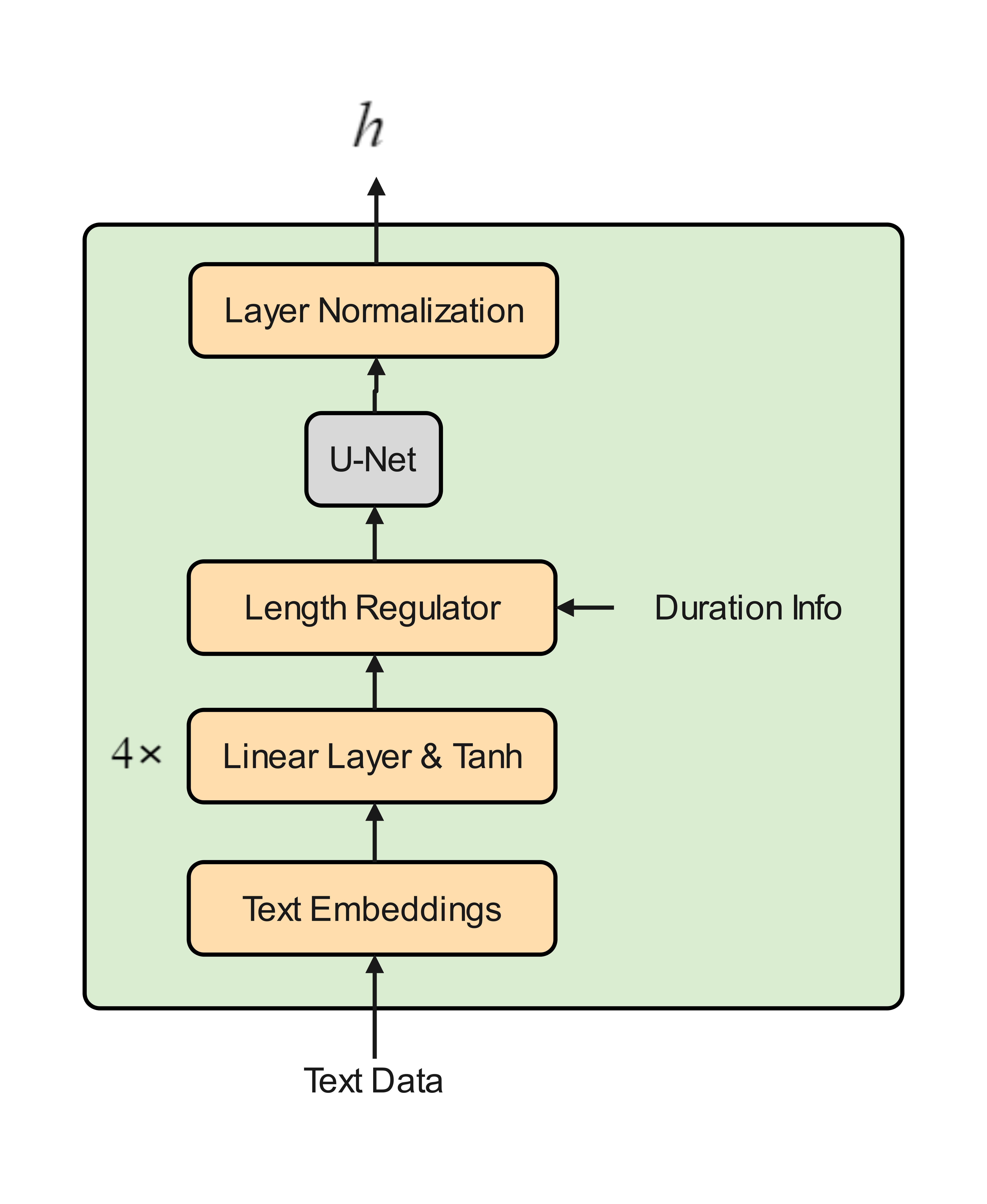}}
        \caption{Overall architecture of block $F$. Here Layer Normalization is performed without re-scaling or re-centering. Re-scaling and Re-centering are afterwards performed in block $G$.}
        \label{fig:2.1}
    \end{subfigure}
    \begin{subfigure}[h]{0.5\textwidth}
  \centering
            \centerline{\includegraphics[scale=0.5]{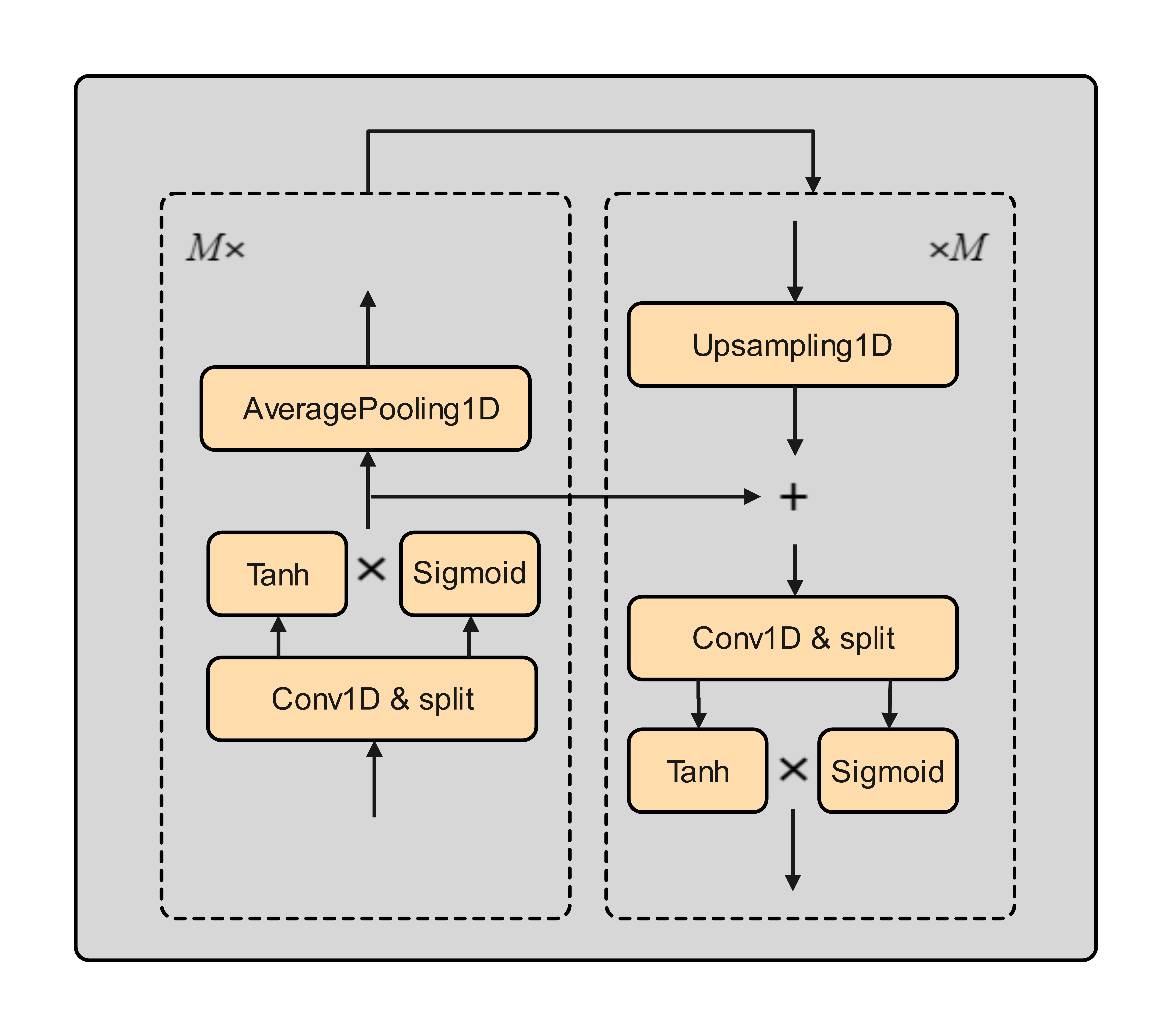}}
        \caption{Detailed architecture of U-Net. $M$ down-sampling blocks (left part) are followed by $M$ up-sampling blocks (right part) with residual connections.}
        \label{fig:U-Net}
    \end{subfigure}
  \caption{Overall structure of block $F$.}
  \label{fig:2}
\end{figure}

To further speed up inference, we use U-Net architecture\cite{U_Net_Convolutional, FPETS} to build block $F$. The down-sampling blocks and up-sampling blocks in U-Net can model long-term dependency while the computation complexity is only linear to sequence length. Following U-Net is LN layer that replaces GRL as discussed in Sec.\ref{subsec:theory}. We use Length
Regulator\cite{fastspeech} that is also used in Fastspeech to deal with length mismatch between text and acoustic feature sequence. The overall architecture of block $F$ is shown by Fig.\ref{fig:2}.



\subsubsection{block $G$}
\begin{figure}[h!]
\centering
\centerline{\includegraphics[scale=0.5]{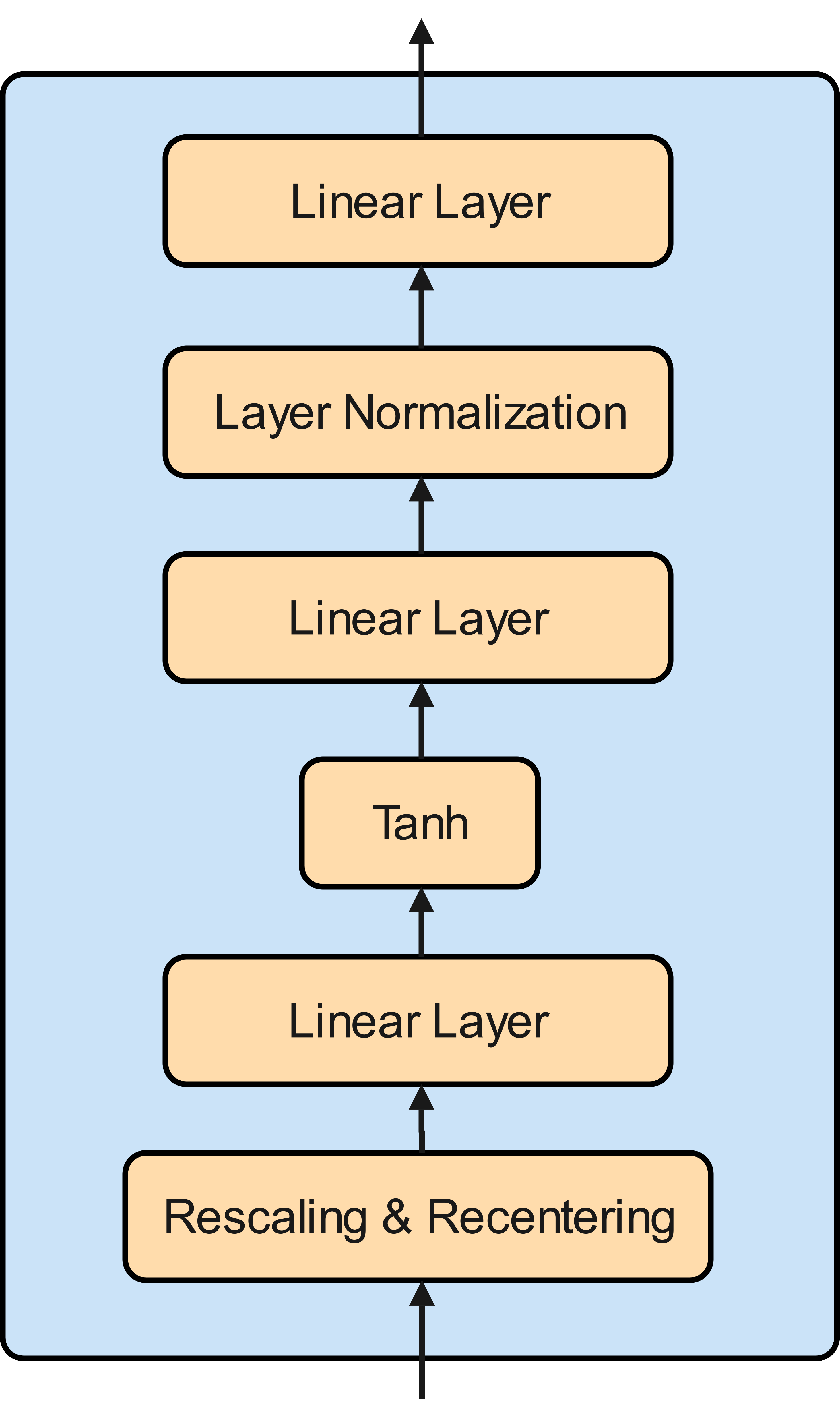}}
\caption{Architecture of block $G$.}\medskip
\label{fig:3}
\end{figure}

As is discussed in Sec.\ref{subsec:theory}, the design of block $G$ is non-trivial and we did experiments on architectures of $G$, with the constraint that temporal connection is not allowed which means there are no convolutional, sequential or fully-connected self-attention layers inside $G$.
Fig.\ref{fig:3} shows the chosen architecture of $G$ which only consists of linear and LN layers.


\section{EXPERIMENTS}

In this section, we test MHTTS on the following subtopics: voice quality and inference speedup in multi-speaker scenario, the ability to handle imperfect transcripts and spontaneous speech in real life. We compare MHTTS with Fastspeech, which is one of the most popular non-autoregressive structure with fast and high-quality voice synthesis.

\subsection{voice quality and inference speedup}
\label{subsec:mutil-speaker}

\subsubsection{experimental setting}
\label{subsubsec:mutil-speaker-experimental-setting}

We conduct multi-speaker experiments on AISHELL-3\cite{AISHELL-3}, a large-scale and high-fidelity Mandarin speech corpus. The corpus contains 85 hours of emotion-neutral 44.1 kHz recordings spoken by 218 native mandarin speakers. We extract mel-spectrogram with 12.5 ms hop size and 50 ms window as acoustic features. A well trained WaveRNN\cite{Wavernn} model is used as vocoder to synthesize voice from mel-spectrogram.

We use the official implementation of Fastspeech\footnote{https://github.com/TensorSpeech/TensorFlowTTS} and the default hyperparameter. We adjust the hyperparameter of MHTTS to use approximately the same number of parameters during inference. The hidden size, kernel size are 512 and 3 respectively for all layers in MHTTS and the number of down-sampling and up-sampling blocks is 7. We use Adam\cite{Adam} optimizer with $\beta_1=0.9,\beta_2=0.98,\epsilon=10^{-6}$ and follow a linear decay learning rate schedule. We train both models for 100K steps.

\subsubsection{evaluation}
We randomly choose 4 speakers from AISHELL-3 speakers(2 men and 2 women, each with 10 utterances) for evaluation. 20 native Mandarin speakers are asked to judge voice quality and similarity of each utterance. Table \ref{lab:1} shows the mean opinion score(MOS) of voice quality, similarity mean opinion score(SMOS), number of active parameters and real time factor(RTF) of 4 methods: GT, the ground truth recording; GT mel + vocoder, voice synthesized with vocoder from mel-spectrogram of ground truth recording; Fastspeech(FS); MHTTS(MH).



\begin{table}[h]
\centering
\resizebox{\columnwidth}{!}{
\begin{tabular}{c|c|c|c|c} 
 \hline
Method & MOS & SMOS & Params & Inference Speed(RTF)\\
 \hline
 GT & $4.05 \pm 0.11$ & $4.40 \pm 0.05$ & - & - \\
 GT mel + vocoder & $3.97 \pm 0.11$ & $4.33 \pm 0.07$ & - & -\\
 FS & $3.53 \pm 0.12$ & $3.78 \pm 0.10$ & 30M & $5.6 \times 10^{-2}$\\
 MH & $3.72 \pm 0.09$ & $3.93 \pm 0.09$ & 27M & $2.3 \times 10^{-2}$\\
 \hline
\end{tabular}}
\caption{MOS, SMOS, number of parameters and inference speed of each case. MOS and SMOS are calculated with 95\% confidence intervals.}
\label{lab:1}
\end{table}

Higher MOS or SMOS means higher quality or similarity. RTF denotes the time (in seconds) required to synthesize one second waveform. It is measured with a single thread and a single core on an Intel(R) Xeon(R) Gold 5218 CPU @ 2.30GHz. It can be seen that in regular multi-speaker scenario, MHTTS achieves slightly higher MOS and SMOS over Fastspeech and achieves 2.4x speedup with the same number of parameters.



\subsection{imperfect transcription}
\label{subsec:imperfect-transcription}
In this section, We conduct experiments to test the ability of transferring text information. We analyze two kinds of imperfect transcription: simulated errors and ASR transcription.

\subsubsection{experimental setting of simulated errors}
\label{subsubsec:experimental setting of simulated errors}
The feature extraction, model structure, and training strategy are identical to Sec.\ref{subsubsec:mutil-speaker-experimental-setting} except the number of training steps is 20K. The dataset is a two-speaker corpora:
\begin{enumerate}
\item \textbf{Source corpus}, a large internal single-speaker Mandarin corpus which contains 140000 female 16 kHz recordings with correct transcription.
\item \textbf{Target corpus}, a well-known open-source high-quality mandarin corpus\footnote{https://www.data-baker.com} which contains 10000 male 48 KHz recordings with correct transcription. All recordings are converted to 16 kHz.
\end{enumerate}

\subsubsection{evaluation of simulated errors}
\label{subsubsec:evaluation of simulated errors}
We corrupt the transcription of the target corpus in a similar way to \cite{investigating_robustness}. 
For every character of a sentence, with probability $P$, the character is corrupted by one of the following corruption methods:
\begin{enumerate}
\item \textbf{Insertion}: a random character from the vocabulary is inserted around it.
\item \textbf{Deletion}: This character is deleted.
\item \textbf{Replacement}: This character is replaced by a random character from the vocabulary.
\end{enumerate}

We vary the corruption probability $P$, the utterance number of the source corpus and the corrupted target corpus and evaluate the character error rate(CER) of the synthesized voice of the target speaker. We compare the following three methods: MHTTS(MH); Fastspeech(FS); Fastspeech\_single(FS\_s), using Fastspeech but training only with corrupted target corpus. For each method we synthesize 200 utterances(2524 characters) for CER evaluation.

\begin{table}[h!]
\centering
\resizebox{\columnwidth}{!}{
\begin{tabular}{c|c|c||ccc} 
 \hline
 \#Source & \#Target & $P$ & FS\_s CER(\%) & FS CER(\%) & MH CER(\%) \\
 \hline
 \multirow{20}{*}{10000} & \multirow{5}{*}{500} & 10\% & 5.62 & 5.07 & 1.03 \\
 
 &&20\% & 9.51 & 5.56 & 0.71 \\
 
 &&30\% & 14.50 & 8.64 & 1.27 \\
 
 &&40\% & 21.16 & 12.04 & 1.03 \\
 
 &&50\% & 32.09 & 14.66 & 1.67 \\
 \cline{2-6}
 & \multirow{5}{*}{1000} & 10\% & 3.24 & 2.46 & 0.79 \\
 
 &&20\% & 6.50 & 2.85 & 0.87 \\

 &&30\% & 9.03 & 4.12 & 1.74 \\

 &&40\% & 15.69 & 7.45 & 1.74 \\
 
 &&50\% & 29.00 & 9.83 & 1.58 \\
 \cline{2-6}
 & \multirow{5}{*}{2000} & 10\% & 3.88 & 1.66 & 0.79 \\
 
 &&20\% & - & 2.06 & 0.87 \\
 
 &&30\% & - & 1.90 & 1.66 \\

 &&40\% & - & 4.04 & 1.90 \\
 
 &&50\% & - & 5.71 & 1.98 \\
 \cline{2-6}
 & \multirow{5}{*}{4000} & 10\% & - & 1.03 & 0.87 \\

 &&20\% & - & 1.74 & 1.11 \\

 &&30\% & - & - & 2.30 \\
 
 &&40\% & - & - & 2.14 \\

 &&50\% & - & - & 1.82 \\
 \hline
\end{tabular}}
\caption{CER evaluated on simulated errors with 10000 utterances of source corpus. "-" indicates that synthesize speeches are unrecognizable sounds.}
\label{table:2}
\end{table}

\begin{table}[h!]
\centering
\resizebox{\columnwidth}{!}{
\begin{tabular}{c|c|c||ccc} 
 \hline
 \#Source & \#Target & $P$ & FS\_s CER(\%) & FS CER(\%) & MH CER(\%) \\
 \hline
 \multirow{20}{*}{50000} & \multirow{5}{*}{500} & 10\% & 5.62 & 3.96 & 0.55 \\

 &&20\% & 9.51 & 4.91 & 0.87 \\
 
 &&30\% & 14.50 & 7.37 & 0.87 \\
 
 &&40\% & 21.16 & 11.65 & 0.95 \\

 &&50\% & 32.09 & 15.53 & 1.66 \\
 \cline{2-6}
 & \multirow{5}{*}{1000} & 10\% & 3.24 & 2.69 & 0.39 \\
 
 &&20\% & 6.50 & 2.38 & 0.79 \\

 &&30\% & 9.03 & 3.57 & 1.35 \\
 
 &&40\% & 15.69 & 8.32 & 1.43 \\
 
 &&50\% & 29.00 & 8.40 & 1.69 \\
 \cline{2-6}
 & \multirow{5}{*}{2000} & 10\% & 3.88 & 1.51 & 0.71 \\

 &&20\% & - & 1.51 & 0.87 \\

 &&30\% & - & 2.22 & 1.03 \\
 
 &&40\% & - & 3.49 & 2.14 \\
 
 &&50\% & - & 4.36 & 1.51 \\
 \cline{2-6}
 & \multirow{5}{*}{4000} & 10\% & - & 1.26 & 1.03 \\

 &&20\% & - & 1.35 & 1.11 \\
 
 &&30\% & - & 1.82 & 1.74 \\
 
 &&40\% & - & - & 1.35 \\

 &&50\% & - & - & 1.74 \\
 \hline
\end{tabular}}
\caption{CER evaluated on simulated errors with 50000 utterances of source corpus.}
\label{table:3}
\end{table}

\begin{table}[h!]
\centering
\resizebox{\columnwidth}{!}{
\begin{tabular}{c|c|c||ccc} 
 \hline
 \#Source & \#Target & $P$ & FS\_s CER(\%) & FS CER(\%) & MH CER(\%) \\
 \hline
 \multirow{20}{*}{100000} & \multirow{5}{*}{500} & 10\% & 5.62 & 3.88 & 0.95 \\
 
 &&20\% & 9.51 & 5.23 & 0.87 \\

 &&30\% & 14.50 & 6.58 & 0.63 \\
 
 &&40\% & 21.16 & 11.73 & 1.27 \\
 
 &&50\% & 32.09 & 13.95 & 1.27 \\
 \cline{2-6}
 & \multirow{5}{*}{1000} & 10\% & 3.24 & 1.66 & 0.48 \\
 
 &&20\% & 6.50 & 3.25 & 0.95 \\
 
 &&30\% & 9.03 & 3.41 & 0.95 \\
 
 &&40\% & 15.69 & 6.26 & 1.82 \\

 &&50\% & 29.00 & 6.66 & 1.35 \\
 \cline{2-6}
 & \multirow{5}{*}{2000} & 10\% & 3.88 & 1.27 & 0.71 \\

 &&20\% & - & 1.51 & 1.03 \\
 
 &&30\% & - & 2.85 & 1.82 \\

 &&40\% & - & 2.22 & 1.98 \\

 &&50\% & - & 4.75 & 1.90 \\
 \cline{2-6}
 & \multirow{5}{*}{4000} & 10\% & - & 1.03 & 0.71 \\

 &&20\% & - & 1.03 & 1.27 \\

 &&30\% & - & 2.06 & 1.82 \\

 &&40\% & - & - & 1.72 \\

 &&50\% & - & - & 1.76 \\
 \hline
\end{tabular}}
\caption{CER evaluated on simulated errors with 100000 utterances of source corpus.}
\label{table:4}
\end{table}

Table \ref{table:2}, Table \ref{table:3} and Table \ref{table:4} show the results and we have several conclusions:

\begin{enumerate}

\item MHTTS performs consistently much better than Fastspeech and Fastspeech\_single, which demonstrates a much better capability of transferring text information.
\item Fastspeech also has the capability of transferring text information when comparing it with Fastspeech\_single.
\item CER of Fastspeech and Fastspeech\_single seem to drop as number of target utterances increases but training collapse may occur. MHTTS performs stably.

\end{enumerate}

\subsubsection{experimental setting of ASR transcription}
\label{subsubsec:experimental setting of ASR transcription}
The feature extraction, model structure, and training strategy are identical to Sec.\ref{subsubsec:experimental setting of simulated errors}. The dataset is a two-speaker corpora:

\begin{enumerate}
\item \textbf{Source corpus}, the same source corpus used in Sec.\ref{subsubsec:experimental setting of simulated errors}. All recordings are converted to 8 kHz.
\item \textbf{Target corpus}, Mandarin spontaneous 8 KHz speeches of 4 speakers(2 men and 2 women) in real life recorded in an internal customer service hotline system. Transcription is not available.
\end{enumerate}

\subsubsection{evaluation of ASR transcription}
\label{subsubsec:evaluation of ASR transcription}
We use an ASR module to transcribe the target corpus, and compare CER of synthesized voice of the target speaker similarly to Sec.\ref{subsubsec:evaluation of simulated errors}. Results are shown in Table.\ref{table:5}. It can be seen that MHTTS performs significantly better than Fastspeech and MHTTS is more robust to ASR transcription errors than simulated errors.

\begin{table}[h!]
\centering
\resizebox{\columnwidth}{!}{
\begin{tabular}{c|c|c|c|c||cc} 
 \hline
 \#Source & ASR CER(\%) & Target ID & Gender & \#Target & FS CER(\%) & MH CER(\%) \\
 \hline
 \multirow{4}{*}{140000} & \multirow{4}{*}{14.09} & A & male & 1164 &9.86 & 0.87 \\

 && B & female & 1265 &6.58 & 0.63 \\
 
 && C & female & 2059 &8.08 & 0.71 \\

 && D & male & 3110 &7.29 & 0.71 \\
 \hline
 
\end{tabular}}
\caption{CER evaluated on spontaneous real data with ASR transcription}
\label{table:5}
\end{table}

\subsection{spontaneous style}
\label{subsec:spontaneous style}

In this section we conduct experiments to test if MHTTS can synthesize expressive voice given spontaneous speech in real life.

\subsubsection{experimental setting}
\label{subsubsec:experimental setting}
The feature extraction, model structure, training strategy and dataset are identical to Sec.\ref{subsubsec:experimental setting of ASR transcription}.

\subsubsection{evaluation}
We use the same ASR module in Sec.\ref{subsubsec:evaluation of ASR transcription} to transcribe the target spontaneous corpus. We compare MOS and SMOS of the following methods: GT, ground truth spontaneous recording; Fastspeech(FS); MHTTS(MH). We ask 20 native Mandarin speakers to judge voice quality and similarity. And they are asked to focus on spontaneous style(timbre, pitch, intonation, rhythm and stress) when evaluating similarity. Table.\ref{table:6} shows the results.

\begin{table}[h!]
\centering
\resizebox{\columnwidth}{!}{
\begin{tabular}{c|c||ccc|ccc} 
 \hline
 \multicolumn{2}{c||}{Settings} &  \multicolumn{3}{c|}{MOS} & \multicolumn{3}{c}{SMOS}\\
 \hline
 $\#Source$ & Target ID & GT  & FS  & MH  & GT  & FS  & MH     \\
 \hline
 \multirow{4}{*}{140000} & A & $4.01 \pm 0.12$ & $2.20 \pm 0.14$ & $3.43 \pm 0.13$ & $4.31 \pm 0.09$ & $2.47 \pm 0.20$ & $3.66 \pm 0.19$ \\

  & B & $3.81 \pm 0.16$ & $2.18 \pm 0.17$ & $3.38 \pm 0.14$ & $4.22 \pm 0.15$ & $2.49 \pm 0.18$ & $3.41 \pm 0.18$\\

 & C & $3.98 \pm 0.13$ & $2.02 \pm 0.17$ & $3.27 \pm 0.18$ & $4.49 \pm 0.11$ & $2.79 \pm 0.20$ & $3.61 \pm 0.18$ \\

 & D & $3.85 \pm 0.17$ & $1.98 \pm 0.16$ & $3.31 \pm 0.17$ & $3.95 \pm 0.13$ & $2.11 \pm 0.18$ & $3.12 \pm 0.19$ \\
 \hline
 
\end{tabular}}
\caption{}
\label{table:6}
\end{table}

As can be seen the performance of Fastspeech is much worse than MHTTS, which is probably caused by two reasons: Fastspeech is not robust to ASR transcription errors; The real spontaneous recordings consist of speeches that contain various pronunciation styles for each phoneme and speeches that contain noises made by the speaker that don't correspond to words. This result demonstrates the effectiveness of MHTTS to deal with spontaneous speeches in real life.

\section{CONCLUSION}
In this paper we propose MHTTS, a fast multi-speaker TTS system that utilizes untranscribed spontaneous speeches in real life. We introduce a multi-head structure that disentangles the learning of text representation and speaker related information, which allows text information transfer and training with spontaneous speeches. We achieve better results in terms of several subtopics compared with a state-of-the-art model. This work is an important step to realize automatic TTS training of spontaneous speeches in real life. For future work, we will explore more details of network structure and training strategy to further reduce CER and synthesize more expressive voice.

\vfill\pagebreak

\bibliographystyle{IEEEbib}
\bibliography{refs}

\begin{thebibliography}{10}

\bibitem{deep_voice2}
Andrew Gibiansky, Sercan Arik, Gregory Diamos, John Miller, Kainan Peng, Wei
  Ping, Jonathan Raiman, and Yanqi Zhou,
\newblock ``Deep voice 2: Multi-speaker neural text-to-speech,''
\newblock in {\em Advances in Neural Information Processing Systems}. 2017,
  vol.~30, Curran Associates, Inc.

\bibitem{deep_voice3}
Wei Ping, Kainan Peng, Andrew Gibiansky, Sercan~Ömer Arik, Ajay Kannan, Sharan
  Narang, Jonathan Raiman, and John Miller,
\newblock ``Deep voice 3: 2000-speaker neural text-to-speech.,''
\newblock {\em CoRR}, vol. abs/1710.07654, 2017.

\bibitem{tacotron2}
R.~J. Weiss M. Schuster N. Jaitly Z. Yang Z. Chen Y. Zhang Y. Wang R.
  Skerrv-Ryan et~al. J.~Shen, R.~Pang,
\newblock ``Natural tts synthesis by conditioning wavenet on mel spectrogram
  predictions,''
\newblock in {\em 2018 IEEE International Conference on Acoustics}. IEEE, 2018,
  p. 4779–4783.

\bibitem{transformertts}
Yanqing Liu Sheng~Zhao Naihan~Li, Shujie~Liu and Ming Liu,
\newblock ``Neural speech synthesis with transformer network,''
\newblock in {\em Proceedings of the AAAI Conference on Artificial
  Intelligence}, 2019, vol.~33, p. 6706–6713.

\bibitem{fastspeech}
X.~Tan T. Qin S. Zhao Z.~Zhao Y.~Ren, Y.~Ruan and T.-Y.Liu,
\newblock ``Fastspeech: Fast, robust and controllable text to speech,''
\newblock in {\em NeurIPS 2019}, November 2019.

\bibitem{fastspeech2}
Yi~Ren, Chenxu Hu, Xu~Tan, Tao Qin, Sheng Zhao, Zhou Zhao, and Tie{-}Yan Liu,
\newblock ``Fastspeech 2: Fast and high-quality end-to-end text to speech,''
\newblock {\em CoRR}, vol. abs/2006.04558, 2020.

\bibitem{ADASPEECH2}
Yuzi Yan, Xu~Tan, Bohan Li, Tao Qin, Sheng Zhao, Yuan Shen, and Tie{-}Yan Liu,
\newblock ``Adaspeech 2: Adaptive text to speech with untranscribed data,''
\newblock in {\em {ICASSP} 2021}. 2021, pp. 6613--6617, {IEEE}.

\bibitem{NAUTILUS}
Hieu-Thi Luong and Junichi Yamagishi,
\newblock ``Nautilus: A versatile voice cloning system,''
\newblock {\em IEEE/ACM Transactions on Audio, Speech, and Language
  Processing}, vol. 28, pp. 2967--2981, 2020.

\bibitem{kaldi}
Daniel Povey, Arnab Ghoshal, Gilles Boulianne, Nagendra Goel, Mirko Hannemann,
  Yanmin Qian, Petr Schwarz, and Georg Stemmer,
\newblock ``The kaldi speech recognition toolkit,''
\newblock in {\em In IEEE 2011 workshop}, 2011.

\bibitem{ContextNet}
Wei Han, Zhengdong Zhang, Yu~Zhang, Jiahui Yu, Chung-Cheng Chiu, James Qin,
  Anmol Gulati, Ruoming Pang, and Yonghui Wu,
\newblock ``Contextnet: Improving convolutional neural networks for automatic
  speech recognition with global context,''
\newblock in {\em interspeech}, 2020.

\bibitem{wenet}
Zhuoyuan Yao, Di~Wu, Xiong Wang, Binbin Zhang, Fan Yu, Chao Yang, Zhendong
  Peng, Xiaoyu Chen, Lei Xie, and Xin Lei,
\newblock ``Wenet: Production oriented streaming and non-streaming end-to-end
  speech recognition toolkit,''
\newblock in {\em interspeech}, 2021.

\bibitem{investigating_robustness}
Jason Fong, Pilar~Oplustil Gallegos, Zack Hodari, and Simon King,
\newblock ``Investigating the robustness of sequence-to-sequence text-to-speech
  models to imperfectly-transcribed training data,''
\newblock in {\em INTERSPEECH}, 2019.

\bibitem{DCTTS}
K.~Uenoyama H.~Tachibana and S.~Aihara,
\newblock ``Efficiently trainable text-to-speech system based on deep
  convolutional networks with guided attention,''
\newblock in {\em ICASSP}, 2018.

\bibitem{transformer}
Ashish Vaswani, Noam Shazeer, Niki Parmar, Jakob Uszkoreit, Llion Jones,
  Aidan~N Gomez, \L~ukasz Kaiser, and Illia Polosukhin,
\newblock ``Attention is all you need,''
\newblock in {\em Advances in Neural Information Processing Systems}. 2017,
  vol.~30, Curran Associates, Inc.

\bibitem{Emergence_of_invariance}
Alessandro Achille and Stefano Soatto,
\newblock ``Emergence of invariance and disentanglement in deep
  representations,''
\newblock {\em J. Mach. Learn. Res.}, vol. 19, pp. 50:1--50:34, 2018.

\bibitem{Unsupervised_Speech_Decomposition}
Kaizhi Qian, Yang Zhang, Shiyu Chang, David Cox, and Mark Hasegawa-Johnson,
\newblock ``Unsupervised speech decomposition via triple information
  bottleneck,''
\newblock in {\em ICML 2020}, 2020, pp. 7792--7802.

\bibitem{Disentangling_Style_Factors}
Jennifer Williams and Simon King,
\newblock ``Disentangling style factors from speaker representations.,''
\newblock in {\em Interspeech}, 2019, pp. 3945--3949.

\bibitem{wavenet}
H.~Zen K. Simonyan O. Vinyals A. Graves N. Kalchbrenner A.~Senior A.~v.~d.
  Oord, S.~Dieleman and K.~Kavukcuoglu,
\newblock ``Wavenet: A generative model for raw audio,''
\newblock in {\em 9th ISCA Speech Synthesis Workshop}, 2016.

\bibitem{wavernn}
K.~Simonyan S. Noury N. Casagrande E. Lockhart F. Stimberg A. van den Oord
  S.~Dieleman N.~Kalchbrenner, E.~Elsen and K.~Kavukcuoglu,
\newblock ``Efficient neural audio synthesis,''
\newblock in {\em ICML}, 2018.

\bibitem{melgan}
Thibault de Boissiere Lucas Gestin Wei Zhen Teoh Jose Sotelo Alexandre de
  Brebisson Yoshua~Bengio Kundan~Kumar, Rithesh~Kumar and Aaron~C Courville,
\newblock ``Melgan: Generative adversarial networks for conditional waveform
  synthesis,''
\newblock in {\em Advances in Neural Information Processing Systems}, 2019, p.
  14881–14892.

\bibitem{information_bottleneck}
Andrew~M Saxe, Yamini Bansal, Joel Dapello, Madhu Advani, Artemy Kolchinsky,
  Brendan~D Tracey, and David~D Cox,
\newblock ``On the information bottleneck theory of deep learning,''
\newblock {\em Journal of Statistical Mechanics: Theory and Experiment}, vol.
  2019, no. 12, pp. 124020, 2019.

\bibitem{Domain_Adversarial}
Yaroslav Ganin, Evgeniya Ustinova, Hana Ajakan, Pascal Germain, Hugo
  Larochelle, Fran{\c{c}}ois Laviolette, Mario March, and Victor Lempitsky,
\newblock ``Domain-adversarial training of neural networks,''
\newblock {\em Journal of Machine Learning Research}, vol. 17, no. 59, pp.
  1--35, 2016.

\bibitem{Layer_Normalization}
Lei~Jimmy Ba, Jamie~Ryan Kiros, and Geoffrey~E. Hinton,
\newblock ``Layer normalization,''
\newblock {\em CoRR}, vol. abs/1607.06450, 2016.

\bibitem{U_Net_Convolutional}
O.~Ronneberger, P.Fischer, and T.~Brox,
\newblock ``U-net: Convolutional networks for biomedical image segmentation,''
\newblock in {\em MICCAI}, 2015, vol. 9351, pp. 234--241.

\bibitem{FPETS}
Dabiao Ma, Zhiba Su, Wenxuan Wang, and Yuhao Lu,
\newblock ``{FPETS:} fully parallel end-to-end text-to-speech system,''
\newblock in {\em {AAAI}}, 2020, pp. 8457--8463.

\bibitem{AISHELL-3}
Yao Shi, Hui Bu, Xin Xu, Shaoji Zhang, and Ming Li,
\newblock ``{AISHELL-3:} {A} multi-speaker mandarin {TTS} corpus and the
  baselines,''
\newblock {\em CoRR}, vol. abs/2010.11567, 2020.

\bibitem{Adam}
Diederik~P. Kingma and Jimmy Ba,
\newblock ``Adam: {A} method for stochastic optimization,''
\newblock in {\em {ICLR} 2015}, 2015.

\end{thebibliography}
\end{document}